\documentclass[twoside]{article}
\usepackage{fleqn,espcrc2}
\usepackage{graphicx}
\usepackage{amsmath}
\usepackage{amsfonts}
\usepackage{slashed}
\usepackage{cite}
\newcommand{\be}{\begin{equation}}
\newcommand{\ee}{\end{equation}}
\newcommand{\bdm}{\begin{displaymath}}
\newcommand{\edm}{\end{displaymath}}
\newcommand{\bea}{\begin{eqnarray}}
\newcommand{\eea}{\end{eqnarray}}
\newcommand{\ba}{\begin{array}}
\newcommand{\ea}{\end{array}}
\newcommand{\nn}{\nonumber}
\newcommand{\pref}[1]{(\ref{#1})}
\newcommand{\chpt}{$\chi$PT}
\newcommand{\oo}[1]{{\cal O}(#1)}

\title{Chiral perturbation theory for lattice QCD at ${\cal O}(a^{2})$\thanks{Talk by  O.~B\"ar at Lattice 2003, Tsukuba.}
}
\author{O. B\"ar\address{Institute of Physics,
        University of Tsukuba, Tsukuba, Ibaraki 305-8571, Japan},
        G. Rupak\address{Lawrence Berkeley National Laboratory,
        Berkeley, CA 94720, U.S.A.} and
        N. Shoresh\address{Department of Physics,
        Boston University, Boston, MA 02215, U.S.A.}
 } 
\begin{document}
\begin{abstract}
The ${\cal O}(a^{2})$ contributions to the chiral effective Lagrangian for lattice QCD with Wilson fermions are constructed. The results are generalized to partially quenched QCD with Wilson fermions as well as to the ``mixed'' lattice theory with Wilson sea quarks and Ginsparg-Wilson valence quarks.
\vspace{0.5pc}
\end{abstract}

\maketitle

\section{Introduction}
Chiral perturbation theory (\chpt) is widely used in analyzing lattice QCD data. Since \chpt\ describes continuum QCD one has to take the continuum limit first before fitting the lattice data to the analytic predictions of \chpt. Unfortunately, lattice data for various lattice spacings is often not available in order to perform a reliable continuum extrapolation. Nevertheless, it is common practice to directly fit lattice data obtained for non-vanishing lattice spacing to continuum \chpt, thus introducing a systematic error.

This uncertainty can be analytically controlled by including the discretization effects stemming from a non-zero lattice spacing in \chpt. For example, the ${\cal O}(a^{2})$ taste violations in staggered fermion simulations have been systematically accounted for that way \cite{Aubin:2003mg}. Fits of MILC simulation data to \chpt\ including ${\cal O}(a^{2})$ taste violating terms show a significantly improved $\chi^{2}$ \cite{ClaudePoster}. In fact, good fits are not possible without these terms.

Similarly, the leading effects linear in $a$ for simulations with Wilson fermions have been included in \chpt\ \cite{Rupak:2002sm} and the resulting expressions for the pseudo scalar mass and decay constant have already been used in fits to simulation data \cite{Farchioni:2003nf}. However, since the lattice spacings in current unquenched simulations with Wilson fermions are usually not very small, it is important to extend the results in \cite{Rupak:2002sm} by including the ${\cal O}(a^{2})$ contributions in \chpt\ \cite{Bar:2003mh,Aoki:2003yv}. 

\section{\chpt\ for Wilson fermions at $\boldsymbol{{\cal O}(a^{2})}$}

The strategy for including the cut-off artifacts in \chpt\ is essentially a two-step matching to effective theories. We first match the lattice theory to Symanzik's effective theory \cite{Symanzik:1983dc}, which is a continuum field theory. This effective theory makes the cut-off dependence of the underlying lattice theory explicit in terms of higher dimensional operators in the effective action and the effective operators. In a second step we construct the chiral effective theory for Symanzik's effective theory, including the extra terms due to the non-zero lattice spacing. The result is a chiral expansion with explicit dependence on the lattice spacing $a$.

In order to include ${\cal O}(a^{2})$ effects in \chpt\ we need Symanzik's effective action for lattice QCD with Wilson fermions through ${\cal O}(a^{2})$.  At this order there are fifteen dimension-6 operators which are compatible with the symmetries of the underlying lattice theory \cite{Sheikholeslami:1985ij}. The way these operators affect the chiral Lagrangian depends on their transformation properties under chiral rotations, and one can basically distinguish three types of operators:
\begin{enumerate}
\item  the O(4) symmetry breaking term\\ $\overline{\psi}\gamma_{\mu}D_{\mu}D_{\mu}D_{\mu}\psi$
\item chiral symmetry conserving terms\\ e.g.\ $\overline{\psi}\slashed{D}^3\psi$ and 
$(\overline{\psi}\gamma_{\mu}\psi)(\overline{\psi}\gamma_{\mu}\psi)$
\item chiral symmetry breaking terms\\  e.g.\ $(\overline{\psi}\gamma_{5}\psi)(\overline{\psi}\gamma_{5}\psi)$
\end{enumerate}
The first operator, which breaks the full O(4) rotation symmetry, is not forbidden since the underlying lattice theory is invariant only under the hypercubic subgroup of O(4). 

The chiral Lagrangian is expanded in powers of momenta $p^{2}$, quark masses $m$ and the lattice spacing $a$. Generalizing the standard chiral power counting, the leading order (LO) chiral Lagrangian contains the terms of $\oo{p^2,m,a}$, while the terms of $\oo{p^4,p^2m,p^2a,m^2,ma,a^2}$ are of next-to-leading order (NLO).  Employing this power counting the first two types of operators given above do not contribute at NLO. 

The first operator, for instance,  gives rise to terms in the chiral Lagrangian which break the O(4) symmetry and are invariant under the hypercubic subgroup only. These terms necessarily contain four derivatives and, since they are also multiplied by $a^{2}$,  are therefore at least of $\oo{p^{4}a^{2}}$. These terms are beyond NLO and can be safely neglected. 

The second type of operators, the chiral symmetry conserving terms, 
modify the coefficients in front of already existing terms in the chiral Lagrangian. For example, the LO coefficient $f^{2}$, multiplying the kinetic term $\mbox{tr}\,\partial_{\mu} \Sigma \partial \Sigma_{\mu} ^{\dagger}$, becomes $f^{2}(a^{2})=f^{2} + a^{2}K+\ldots$, with some other (unknown) coefficient $K$. This leads to the extra term $a^{2}K\mbox{tr}\,\partial_{\mu} \Sigma \partial \Sigma_{\mu} ^{\dagger}$ in the chiral Lagrangian, which is of $\oo{a^2p^2}$ and thus beyond NLO. This example can be generalized to all chiral symmetry conserving terms and we conclude that these terms do not affect the NLO chiral Lagrangian.

Only some of
the chiral symmetry breaking terms give rise to new operators in the NLO chiral Lagrangian. The way they influence the chiral Lagrangian is obtained by a standard spurion analysis and the final result for the contributions proportional to $a^{2}$ reads
\bea\label{Lag1}
{\cal L}[a^{2}] & = & - \hat a^2 W_6^{'}\big\langle\Sigma^\dag+\Sigma\big\rangle^2
-\hat a^2W'_7\big \langle\Sigma^\dag-\Sigma\big\rangle^2\nonumber\\[0.2ex]
& & -\hat a^2W'_8\big\langle\Sigma^\dag\Sigma^\dag+\Sigma\Sigma\big\rangle.
\eea
Here $\Sigma$ is the standard matrix of Nambu-Goldstone fields and the angled brackets denote traces over the flavor indices. The parameter $\hat a = a W_{0}$ absorbs the leading order low-energy constant $W_{0}$ in order to render the new unknown low-energy constants $W_6^{'},\,W_7^{'},\, W_8^{'}$ dimensionless. Note that only three new operators appear at ${\cal O}(a^{2})$ despite the fact that there are many different operators in the underlying Symanzik action at this order.

We are now in the position to compute NLO expressions for physical observables. For example, the ${\cal{O}}(a^{2})$ contribution to the flavor non-singlet pseudo-scalar mass is given by 
\be\label{deltaM}
\delta M_{PS}^{2}[a^2]\, = \,\frac{16}{f^{2}} \,\hat a^2 ( N_{f } W_6^{'} + W_8^{'}),
\ee
where $N_{f}$ denotes the number of flavors. Adding this contribution to the terms of ${\cal O}(a)$ given in \cite{Rupak:2002sm} we obtain the complete NLO result through ${\cal O}(a^{2})$ for the pseudo-scalar mass. 

\section{Partially quenched QCD}
The previous discussion is readily extended to partially quenched lattice QCD with different masses for the sea and valence Wilson fermions. Described by a lattice action with sea, valence and ghost quarks, the Symanzik action through ${\cal O}(a^{2})$ for  partially quenched lattice QCD is obtained as in the unquenched case, based on locality and the symmetries of the lattice theory. The result for the ${\cal O}(a^{2})$ contribution to the partially quenched chiral Lagrangian has the same form as in \pref{Lag1}, with the angled brackets now denoting super-traces and the field $\Sigma$ reflecting the larger flavor content of partially quenched \chpt.

The discussion of  ``mixed fermion theories'' is slightly more complicated. Mixed theories are a generalization of partially quenched lattice theories. In addition to choosing different quark masses, the lattice Dirac operator is different in the sea and valence sector. The concrete example  
with the Wilson Dirac operator for the sea quarks and a Dirac operator satisfying the Ginsparg-Wilson relation for the valence quarks has been proposed in \cite{Bar:2002nr}. 

Such a combination has various advantages. As a consequence of the Ginsparg-Wilson relation the valence sector exhibits an exact chiral symmetry and therefore does not suffer from the shortcomings of Wilson fermions due to their explicit chiral symmetry breaking (additive mass renormalization, cut-off effects linear in $a$, etc.). In particular, simulations are possible with valence quark masses much smaller than accessible with Wilson fermions \cite{hasenfratz}. This should allow numerical simulations deeper in in the chiral regime of partially quenched QCD.
Finally, mixed fermion simulations offer a cost-effective compromise towards full unquenched simulations using Ginsparg-Wilson fermions, since mixed simulations only require the computation of correlators in the background of already existing unquenched configurations generated with Wilson fermions.

Deriving the ${\cal O}(a^{2})$ corrections in the chiral Lagrangian for the mixed fermion theory with Wilson sea and Ginsparg-Wilson valence quarks goes along the lines outlined above. However, due to the use of Ginsparg-Wilson fermions for the valence quarks there are no  ${\cal O}(a)$ terms and fewer ${\cal O}(a^{2})$ terms in the valence sector of Symanzik's effective action for the mixed theory. As before one can show that the O(4) symmetry breaking terms as well  as the chiral symmetry conserving operators in Symanzik's action do not contribute to the NLO chiral Lagrangian. A spurion analysis for the chiral symmetry breaking terms gives the following
${\cal O}(a^{2})$ contribution for the chiral Lagrangian:
\bea\label{Lag2}
{\cal L}[a^{2}] & = & - \hat a^2 W_6^{'}\big\langle P_{S}\Sigma^\dag+\Sigma P_{S}\big\rangle^2\nn\\
& &-\hat a^2W'_7\big \langle P_{S}\Sigma^\dag-\Sigma P_{S}\big\rangle^2\nn\\
& & -\hat a^2W'_8\big\langle P_{S}\Sigma^\dag P_{S}\Sigma^\dag+\Sigma P_{S}\Sigma P_{S}\big\rangle\nn\\
& & -\hat a^2W_M\big\langle\tau_{3}\Sigma\tau_{3}\Sigma^{\dagger}\big\rangle
\eea
The operators $P_{S}$ and $P_{V}$ denote projection operators to the sea and valence sector, respectively, and $\tau_{3}$ equals $P_{S} - P_{V}$ (as before,  $\hat a = a W_{0}$).

The first three terms stem entirely from the Wilson fermion sea sector. They involve sea sector fields only (note the projector $P_{S}$) and they are the same three operators as in \pref{Lag1}. The additional operator mixes the sea and valence sector. The reason for the presence of the extra term is the reduced flavor symmetry group of the mixed theory. The use of different Dirac operators for sea and valence quarks does not contain transformations between sea and valence fields and the additional term $\langle\tau_{3}\Sigma\tau_{3}\Sigma^{\dagger}\rangle$ is not forbidden by the symmetries of the theory.

The presence of this extra term in the chiral Lagrangian does not  entail that chiral expressions for observables in the mixed theory depend on more free parameters. For example, taking account of the terms in \pref{Lag2} in the calculation of the flavor non-singlet pseudo-scalar mass one finds that its ${\cal O}(a^{2})$ correction actually vanishes. 
This is not that surprising. Exact chiral symmetry in the massless case for Ginsparg-Wilson
fermions implies that the pseudo-scalar meson mass is proportional to the quark
mass and vanishes in the chiral limit. Hence any lattice spacing correction to $M_{PS}^2$ is suppressed by  at least one factor of the valence quark mass $m_{Val}$. The largest lattice correction quadratic in the lattice spacing is therefore of $\oo{m_{Val}a^2}$. This example illustrates that certain beneficial properties of Ginsparg-Wilson fermions are preserved even in the presence of a ``non-Ginsparg-Wilson'' sea sector. 



\begin{thebibliography}{20}

\bibitem{Aubin:2003mg}
C.~Aubin and C.~Bernard,
Phys.\ Rev.\ D {\bf 68}, 034014 (2003) and arXiv:hep-lat/0306026.



\bibitem{ClaudePoster}
C.~Aubin {\em et.\ al.,}
these proceedings (poster presentation by C.\ Bernard).


\bibitem{Rupak:2002sm}
G.~Rupak and N.~Shoresh,
Phys.\ Rev.\ D {\bf 66}, 054503 (2002).

\bibitem{Farchioni:2003nf}
F.~Farchioni, I.~Montvay, E.~Scholz and L.~Scorzato  [qq+q Collaboration],
arXiv:hep-lat/0307002.

\bibitem{Bar:2003mh}
O.~B\"ar, G.~Rupak and N.~Shoresh,
arXiv:hep-lat/0306021.

\bibitem{Aoki:2003yv}
S.~Aoki,
arXiv:hep-lat/0306027.

\bibitem{Symanzik:1983dc}
K.~Symanzik,
Nucl.\ Phys.\ B {\bf 226}, 187 (1983) and Nucl.\ Phys.\ B {\bf 226}, 205 (1983).



\bibitem{Sheikholeslami:1985ij}
B.~Sheikholeslami and R.~Wohlert,
Nucl.\ Phys.\ B {\bf 259}, 572 (1985).

\bibitem{Bar:2002nr}
O.~B\"ar, G.~Rupak and N.~Shoresh,
Phys.\ Rev.\ D {\bf 67}, 114505 (2003).


\bibitem{hasenfratz}
see, e.g., the talks by K.~F.~Liu and P.~Hasenfratz, these proceedings.

\end{thebibliography}
\end{document}